\documentclass[12pt,preprint]{aastex}
\slugcomment{Accepted for publication 
in {\it the Astrophysical Journal Letters}  July 8, 2003} 
\begin{document}


\title{The Bursting Behavior of 4U 1728-34: Parameters of a Neutron Star and 
Geometry of a  NS-disk system}

\author{Nickolai Shaposhnikov\altaffilmark{1}, 
        Lev Titarchuk\altaffilmark{1,2} and 
	Frank Haberl\altaffilmark{3}}

\altaffiltext{1}{George Mason University, 
Center for Earth Observing and Space Research, Fairfax, VA 22030; 
nshaposh@scs.gmu.edu; lev@xip.nrl.navy.mil}
\altaffiltext{2}{NASA/ Goddard Space Flight Center, Greenbelt 
MD 20771, USA; lev@lheapop.gsfc.nasa.gov}
\altaffiltext{3}{Max-Planck-Institut f\"ur extraterrestrische Physik,
Giessenbachstra{\ss}e, 85748 Garching, Germany, fwh@mpe.mpg.de}

\begin{abstract}
We analyze a set of Type I X-ray bursts from the low mass X-ray binary 4U 1728-34, 
observed with {\it Rossi X-ray Timing Explorer} (RXTE).
We infer the dependence of the neutron star (NS) 
mass and radius with respect to the assumed distance to the system using 
an analytical model of X-ray burst spectral formation.
The model behavior clearly indicates that the burster atmosphere is 
helium-dominated. Our results strongly favor the soft equation of
state (EOS) of NS  for 4U 1728-34. We find that distance to the source should be
within 4.5-5.0 kpc range. We obtain rather narrow constrains for the NS radius in 8.7-9.7 km range 
and interval 1.2-1.6 $M_\odot$ for NS mass for this particular distance range. 
 We uncover 
 a temporal behavior of red-shift corrected burst flux  for the radial 
expansion episodes and we put
 forth 
a dynamical evolution scenario for the
NS--accretion disk geometry during which an expanded envelope affects the  accretion
disk and increases the area of the neutron star  exposed to the Earth observer. 
In the framework of this scenario we provide a new method for the estimation
 of the inclination angle which leads to the value of
$\sim 50^\circ$ for 4U 1728-34 .

\end{abstract}
\keywords{accretion, accretion disks---stars:fundamental parameters---stars:individual(4U 1728-34)
--- X-ray: bursts}

\section{Introduction}
 A low mass X-ray binary (LMXB) system 
 consists of a neutron star (NS) which accretes matter through Roche lobe overflow 
from an evolved low-mass secondary star. 4U 1728-34 has been recognized as a classical
LMXB because it exhibits a wide range of observational characteristics
 which are attributed to LMXBs [see e. g. \citet{lpt}].
In particular, it exhibits regular thermonuclear
explosions of accreted matter on the NS surface  [Type I X-ray 
bursts, see \citet{str2003} for the latest review]. 
High-frequency quasi-periodic oscillations (kHz QPO)
in persistent emission [\citet{fv,mk}], 
and 363 Hz burst oscillation 
[\citet{str}, \citet{franco},  \citet{S01}, hereafter VS01] were revealed 
 in burst emission from 4U 1728-34 using Fourier analysis 
of the high time-resolution RXTE data.  
So far, no optical counterpart for this X-ray source has  been found. 
Although 4U 1728-34 is believed to be located several kiloparsecs 
from the Earth, no accurate estimation of the distance to the source is currently
available.
The theory of X-ray spectral formation during the expansion and contraction
stages of the bursts   was developed 
in \citet{t94a} and \citet{burst},  hereafter T94 and ST02 respectively.
This theory was first applied to EXOSAT data in \citet{ht}, hereafter HT95,
 for the LMXBs
4U 1705-44 and 4U 1820-30. In \citet{cygx2}, hereafter TS02, three bursts from
Cyg X-2 were analyzed. 
  In this work we  employ the
 methodology, developed in TS02  to analyze a set of 26 bursts from 4U 
1728-34, previously analyzed in VS01 to  search for burst oscillations.
Compared with Cyg X-2 (TS02), 
the 4U 1728-34 burst data have the advantage of high quality 
counting statistics as well as a larger number of burst events.

A brief description of the data  used in the analysis is given 
in \S 2. We present the model 
and the results of its application to the burst data of 4U 1728-34 
in \S 3. Specifically, we obtain the dependence of the NS mass on the radius 
as error contours, calculated for the set of distances to the system taken 
from reasonable interval. 
In \S 4
we offer  an evolution scenario for the NS - accretion disk geometry, 
which can explain the existing controversy between the Eddington limit for
 the peak flux and the flux behavior during the bursts with radial expansion
 \citep{str2003}. In \S 4 we also present estimates for the inclination angle 
of the system. 
We discuss  our results and come to conclusions~in~\S5.

\section{Observations} 
We analyzed the data collected by the Proportional  Counter Array,
\citep[PCA;][]{pca}, 
the main instrument on 
board the {\it RXTE}. Generous amount  ($>$ 1100 ksec) of {\it RXTE} 
observational time 
was devoted to 4U 1728-34. More than 70 bursts were detected.  
For our analysis we selected 
 the subset of bursts based on statistical equivalence and PCA data 
configuration homogeneity, namely, 
when all five detectors of PCA are 
operational and detailed spectral analysis on a subsecond time scale is
possible. The selected events occurred during three periods: 15 March - 
1 February 1996  (Proposal ID 10073), 19 September - 1 October 1997 (Proposal ID 20083),
and 28  February - 10 June 1999 (Proposal ID 40027).
 A total of 26 Type I bursts were selected. 
Due to its exotic nature we excluded the second 
burst detected on 26 September 1997 (Observation ID 20083-01-04-00,
burst 19  according VS01 numbering).

\section{Data analysis and results}

We extract spectral
slices from the Burst
Catcher Mode for consecutive 0.125 sec time intervals for each burst.  
We  obtain the spectrum of the persistent emission using a 300-500 
second time interval prior to a particular burst
and  we input resulting 
spectra as a background to distinguish the burst radiation component. 
 We use fixed hydrogen column of $N_H=1.6\times 10^{22}$ provided 
by HEASARC\footnote{ http://heasarc.gsfc.nasa.gov/cgi-bin/Tools/w3nh/w3nh.pl} 
to model Galactic absorption.
The dead-time corrections is applied to all spectra.
We fit
the burst emission component using a blackbody model. This is justified because
the X-ray burst spectrum deviates from the blackbody-like shape
only in the soft part $\lesssim 1$ keV (T94 and ST02).
The 
quality of the fits is quite good for all spectra except the particular contraction episodes 
when the luminosity  is very close to the Eddington and the photospheric radius changes rapidly along
with  the outgoing spectrum shape.
We calculate model flux between 0.01 and 100 keV. 
Errors for parameter estimations from spectral fits are calculated  
for 68\% (1$\sigma$) confidence level.  
  For the interpretation of the spectral fit results we utilized  the
theoretical model
for the  color temperature of the spectra from the burst cooling phase.
The underlying formalism is developed in T94 and TS02. Here we present the
final formula for the color temperature $kT_{\infty}$(see Eqs. 7-8 in
TS02) 
\begin{equation}
\label{kT}
kT_{\infty}=2.1\, T_h\{l
m[(2-Y_{He})(z+1)^3r_6^2]^{-1}\}^{1/4}{\rm keV}
\end{equation}
where $m$ is the NS mass in units of solar mass, $r_6$ is the NS radius in
units of 
10 km, $Y_{He}$ is the helium abundance, $l=L/L_{Edd}$ is the
dimensionless luminosity in
units of the Eddington luminosity. $T_h$ is the color (hardening) factor,
which  depends on $l$, and $Y_{He}$  (TS02). 
Parameters of the model are $m$,
$r_6$, $Y_{He}$ and $d_{10}$ - distance to the object in units of 10
kpc. The dimensionless luminosity (Eddington ratio) is expressed by 
\begin{equation}
\label{dimlum}
l=0.476\,\xi_b\,d_{10}^2 F_8 (2-Y_{He}) (z+1)/m,
\end{equation}
where $\xi_b$ is an anisotropy factor, $F_8=F/(10^{-8}$ 
erg~cm$^{-2}$~s$^{-1}$) is the observed bolometric flux.
Equations (\ref{kT}) and (\ref{dimlum}) describe the
functional dependence of the observed $kT_{\infty}$ on
the observed flux $F_8$ for the set of
input parameters $m$, $r_6$, $Y_{He}$ and $d_{10}$.
In the next section we put forth the NS-accretion disk geometry scenario
and infer the behavior of $\xi_b$ during a burst with radial expansion. 
In this  proposed scenario, 
the transition from the expansion stage to the
decay stage  is  described graphically 
on the upper panel of Figure \ref{geometry}.
At the beginning of the decay stage, 
immediately after the expansion stage ends, the entire star 
is exposed to the observer.
We do not need any correction  due to the system geometry, which means that
$\xi_b=1$ in this case. Then at some moment $t=t^*$ accretion disk recedes
(reaches the star surface) and  a certain part of the NS is obscured by 
the disk for the observer. The anisotropy factor $\xi_b^*$ that  quantitatively 
takes into account this occultation effect is more than one in this case.
In terms of  functional dependence of $kT_\infty$ upon $F_8$ the flux domain 
 consists of three intervals. For $F_8>\xi_b^*F_8^*$ the entire star
is open and $\xi_b=1$. The flux level $\xi_b^*F_8^*$ from the NS surface 
is related to the star-disk position  when the lower NS hemisphere   starts to get  covered 
by the disk. 
Through the decay stage  the temperature should be calculated  using formulae 
(\ref{kT}) and (\ref{dimlum}) where  $\xi_b=\xi_b^*>1$.
Thus  the model formulated in this way acquires  
two more parameters, the anisotropy factor during occultation $\xi_b^*$
and dimensionless flux $F^*_8$ at the moment when occultation occurs.
Our analysis shows that constant  values of these parameters  
for all bursts with radial expansion are consistent with observations.
We use $\xi_b^* = 1.2$ 
that is estimated using the expansion stage evolution (see \S~4). 
The best-fit value for $F_8^*=5.7$ does not significantly depend on the
distance either.

We combine fluxes and temperatures for all 26 bursts and fit the model 
to the the entire  data set in the range $1.0 < F_8 < 9.5$, which
approximately corresponds to $0.1 < l < 0.9$. We use the 
lower limit to exclude the data points with large errors. We put  the
upper  limit on $l$ because of the restricted validity of the model for $l\sim 1$
where expansion of the atmosphere can occur. We  use a set of  values 
for the distance between 4 and 5 kpc. 
We found that  our analytical model gives 
statistically acceptable fit only when $Y_{He}>0.9$.  In fact, for lower $Y_{He}$ 
 the slope of the model color temperature  dependence on $F_8$ is steeper than that is 
 dictated by the data (see Fig. 2). The model with hydrogen-reach
gas composition fails to describe the data, particularly, due to 
the strong dependence of the hardening factor on the Eddington ratio $l$.
A given change in flux represents a larger change in the Eddington ratio in an $H$-atmosphere than that 
in an $He$-atmosphere because the value of $L_{Edd}$ is smaller for an $H$-atmosphere.
We consider this fact  as a strong argument  for NS atmosphere in 4U
1728-34 to be helium dominated and further we assume $Y_{He}=1.0$ during the whole
proceeding analysis. In Figure 2 we present the data and the best-fit model  for the case of
$d_{10}=0.45$  for which we obtain $M_{NS}=1.25^{+0.06}_{-0.04}\,M_\odot$ and
$R_{NS}=9.00^{+0.17}_{-0.28}$ km.
The best-fit values of NS mass and radius and error 
contours for 68\%,  
90\% 
 and 99\%
confidence levels are  obtained for each fit and they are presented
in Table 1 and Figure \ref{contours}.   
The number of degrees of freedom is 1418 and thus the acceptable fits must 
satisfy the condition that  $\chi^2_{red}=\chi^2/1418\leq 1.0$.  
This condition is not satisfied  for distances higher than  5 kpc for 
which $\chi^2_{red}$ grows very rapidly. 
This fact suggests that probability for NS mass to be higher than 1.6 $M_{\odot}$ is  very low.
The dashed line in Figure \ref{contours} presents the 
dependence of the inner disk radius $R_{in}$ on the NS mass $m$, 
derived for 4U 1728-34 using 
the transition layer model (TLM) for QPOs detected in the persistent state
from LMXBs
 \citep{to,lib}: $R_{in}=9\times m^{1/3}~{\rm km}$.
 Our values of the NS radius  are in  good agreement
with the TLM. Based both on this fact and on statistical
behavior of the model we can conclude {\it that 4.5 - 5.0 kpc is the most probable 
interval for the distance to the source, that  relates to 8.7-9.7 km and
1.2-1.6 $M_\odot$ ranges  for $R_{NS}$ and $M_{NS}$ respectively.}
 
\section{Dynamic evolution of the system geometry during the expansion stage}

\begin{figure}
\includegraphics[width=6.5in,height=6.5in]{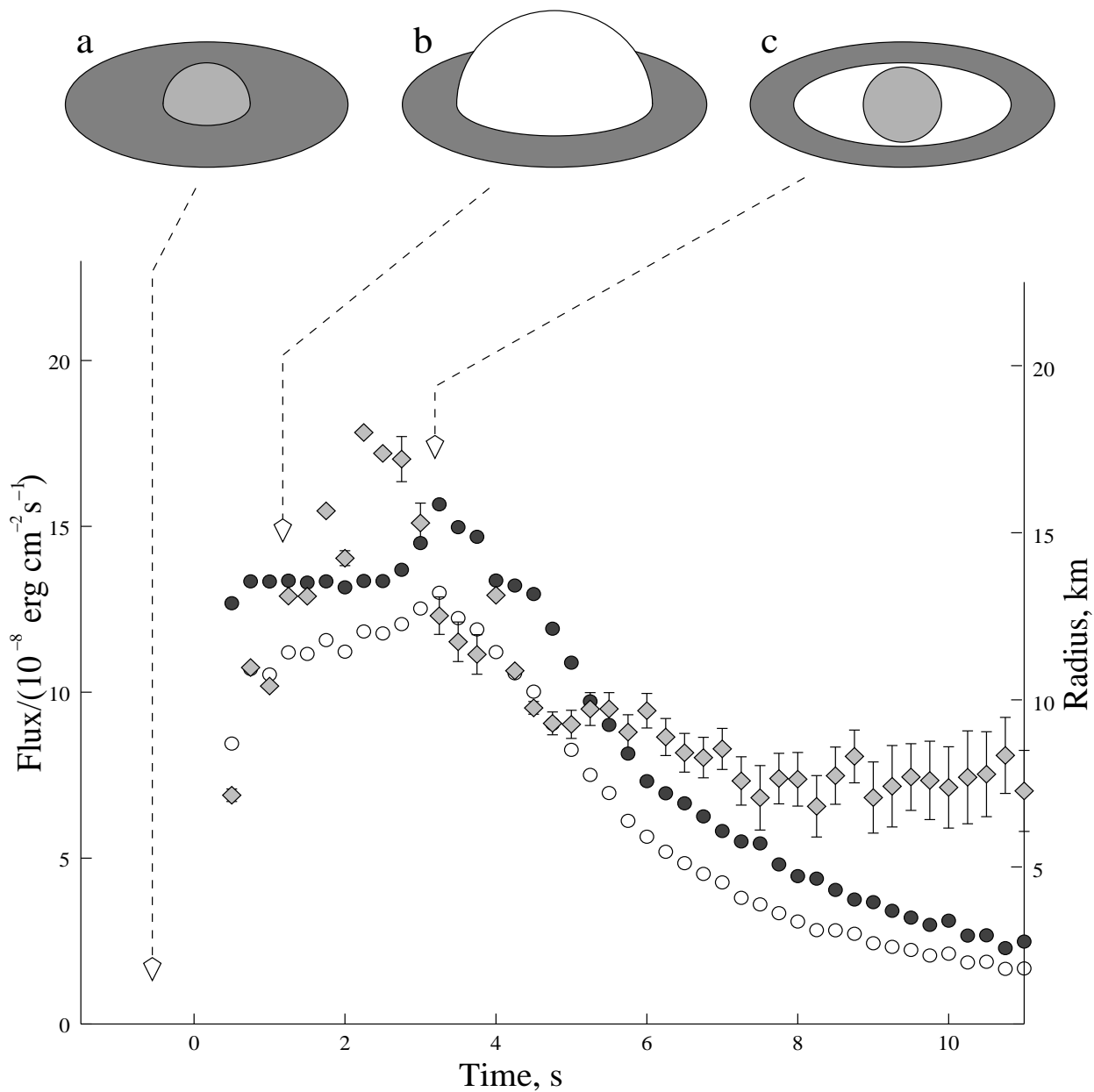}
\caption{ Geometry evolution of the burst through radial expansion. 
White  circles present observed bolometric flux. Filled circles present
GR corrected flux. 
Photospheric radii inferred for the  pure helium 
atmosphere and distance of 4.5 kpc are shown by diamonds. 
Upper panel displays a cartoon diagram of different states of a NS-accretion disk system.  
Dashed arrows point to different stages of the burst.\label{geometry}}
\end{figure}

We investigate  the temporal evolution  of the burst atmosphere
 photospheric radius for each individual burst in the manner 
similar to TS02 and HT95. 
Choosing particular values of distance and NS mass (obtained from the model fit) 
we find the radius [i.e. solve equation (7) of TS02] for each spectral slice. 
The evolution of the NS photospheric radius during the burst event 
is shown in Figure \ref{geometry} for the case of $d_{10}=0.45$. 
We present  radius (diamonds) and observed flux (empty circles)  values 
 versus time for
 burst 9, according to the VS01 numbering  convention (observation ID
 10073-01-06-00). The data were rebinned with 0.25 second time resolution for
presentation purpose. Filled circles represent the  red-shift-corrected  flux  
for each data point. This  red-shift recalculated flux,
 is  the flux which would be detected by an observer situated on  the NS 
photosphere.
Restoration of the red-shift corrected flux reveals distinctive
features which are barely seen  in the observed flux behavior.
After the initial rise of the burst,  the flux levels off and stays constant
for more than a second while the  radius increases gradually. After the
 radius reaches its maximum, a second rise in the flux occurs.
The flux reaches its maximum value when radius begins to fall
quickly in contrast to the initial flux plateau.  After that,
flux decays exponentially, indicating the end
of the expansion stage, and radius levels off.

The above analysis clearly indicates that the flux emitted in the
direction of the  Earth, (when measured locally at the NS surface), is not
constant throughout the expansion stage. This behavior of the burst peak flux
was found to be common for many bursters [see  VS01, \citet{gall}]. 
If the
system geometry remains unchanged,  the mentioned flux behavior 
is in contradiction with the 
Eddington limit for the radiation power from  a stellar atmosphere. 
We argue that the expanded burst envelope interacts strongly with
the inner accretion disk. The evolution of 
the system geometry through the burst event  is displayed in the upper
part of Figure \ref{geometry}. Before the burst the accretion disk extends
down to the surface of NS (stage a), 
{\it covering the lower part of the NS, which is not exposed to the  
Earth observer}. 
Then the burst starts and the atmosphere expands (stage b). At this moment, the
 inner part of the disk is swept away by the burst radiation pressure 
in excess  of Eddington flux. This stage 
corresponds to the initial plateau of the red-shift corrected flux versus 
time.  After the 
photosphere begins to contract, the second rise in the  red-shift corrected
 flux starts, effectively indicating that {\it the lower NS hemisphere appears 
 from behind  the accretion disk}.
Indeed, the free fall velocity is much
higher than the radial propagation velocity component in the disk. After the 
touchdown of the burst envelope,  the NS-disk configuration corresponds 
to the case (c). We assume that the red-shift corrected flux obtained 
for geometry (b) is  the  critical (upper limit)
hemisphere flux.
 Obviously, the red-shift corrections  depend on the NS mass and radius.  
 Using
the standard disk theory \citep{ss} one can estimate the expansion stage duration required for 
sufficient disk material evacuation  as
\begin{equation}
{\cal T}_{Exp} \approx 10^{-6} m/(\alpha \dot{m}^2 \varepsilon)~~{s},
\end{equation}
where $\alpha$ is efficiency of the radial momentum transfer, $\dot{m}$ 
is the mass accretion rate in units of critical mass flux value and $\varepsilon\approx 0.03 \sim 0.05$ 
is a burst flux fraction transformed into the potential energy of 
ambient gas (see e.g. ST02).
For burst sources 
the persistent mass accretion rate
in the disk  is believed to be  $\dot{m}=0.1\sim 1$ [see e. g. \citet{lpt}]. 
Values for  $\alpha\sim 0.1$ are widely used in the
astrophysical community for LMXB.
Under these  assumptions  it  takes only a small fraction of a second 
(${\cal T}_{Exp}\ll 0.1$ s)
to push the inner disk edge out, while observed 
expansion episodes of strong bursts from 4U 1728-34 usually last more than a 
second.
This simple estimate suggests that the expanded NS atmosphere 
effectively push accretion disk outward.
 
 As long as the total NS luminosity during expansion stage (the Eddington luminosity) is constant 
the  observed radiation flux 
is higher 
in geometry (c) than in (a) and (b). Assuming that the NS
radiates at the Eddington limit, the ratio of fluxes detected from the direction   
 at inclination  angle $i$  from the normal to the accretion disk in geometries (b) and (c)
is
\begin{equation}
\tilde{F}_b/\tilde{F}_c=H(i)/H(0),\;~~~ H(i)=\int_{i-\pi/2}^{\pi/2}\cos \omega d\omega \int_{-\pi/2}^{\pi/2}
I(\mu)\cos^2 \psi d\psi,
\end{equation} 
where $\omega$ and $\psi$ are starcentric coordinates, 
$\mu=\cos \omega \cos \psi$ and $I(\mu)$ describes the angular intensity distribution law (see Sobolev 1975 for
details). The tilde denotes the fact that value of flux was corrected for redshift.  For the burst, presented in 
Figure \ref{geometry}, we have $\tilde{F}_b\simeq 1.3\times 10^{-7}$ erg/cm$^2$ and 
$\tilde{F}_c\simeq 1.55 \times 10^{-7}$ erg/cm$^2$. Assuming $I(\mu)=$ const  
(the Lambert law), we obtain \mbox{$\tilde{F}_b/\tilde{F}_c=(1+\cos i)/2$} which leads to an estimate of 
the inclination angle  $i\sim 50^\circ$. This result is also close to that
for the Chandrasekhar angular distribution, $I=I_0(1+2.06\mu$).
We can apply  the value of $\tilde F_c$ for the quantitative assessment 
of the distance to the source.  The peak of the red-shift corrected flux $\tilde F_c$ corresponds 
to  the state when the burst atmosphere subsides on
the NS surface. 
The anisotropy factor is estimated as
$\xi_b=\tilde{F}_{c}/\tilde{F}_b\simeq 1.2$.
We use this value as the anisotropy factor $\xi_b^*$ in the a-geometry
.

\section{Discussion and Conclusions}

In this {\it Letter} we offer a sophisticated
analysis  of the burst expansion events. Accounting for the general relativistic (GR) effects 
reveals the  dynamics and geometry
of the NS-disk system and consistently explains the controversy
between theory and the behavior observed during the  expansion episodes 
of X-ray bursts.
As a direct outcome of the proposed geometric scenario
we obtain the distance   and estimate the inclination angle of the system.
The derived mass-radius relation depends on
the assumed anisotropy. For example, without taking into account of anisotropy, $\xi_b=1$ (conventual
approach) we obtain $M/M_{\odot}=1.29$ and $r=7.8$ km for $d=4.5$ kpc while accurate anisotropy corrections 
implemented in  \S 3 result in  $M/M_{\odot}=1.25$ and $r=9.0$ km.

The statistical behavior of our model clearly rules out values 
of helium abundance lower than 0.9. We could not find the set of
model parameters including the distance that would give an acceptable
value of $\chi^2$ for $Y_{He}<0.9$. The lower values of helium abundance
gives much steeper temperature versus flux functional dependence 
than it is dictated by the  data.  
Furthermore, in earlier observations of 4U 1728-34  Basinska et al. (1984)
 found that 
 helium was probably the main
 source of nuclear fuel for the bursts and that any contribution from
hydrogen was small.  Our data analysis  confirms the results of Basinska et al. and gives another argument 
for a helium-reach atmosphere in 4U 1728-34.
 Our results also favor  
the soft EOSs [\citet{bp}].

We conclude that  (1) the application of  our  analytical model 
to the data leads to the determination of the NS mass and radius 
 as a function of the distance to the system. For the range of allowed distances 4.5-5.0 kpc we obtain
rather narrow constrains for the NS radius in 8.7-9.7 km range and
wide interval 1.2-1.6 $M_\odot$ for NS mass (see also discussion in TS02 and 
Strohmayer \& Bildsten 2003); 
 (2) 
the consistent evolution scenario for the NS - accretion disk geometry,
which explains the variation of the peak flux during the radial expansion stage;
(3) our geometrical model enables us to  estimate;
the inclination angle of  the system to $i\sim 50^\circ$ with respect to the Earth observer.

We acknowledge the fruitful and constructive discussion with the referee. 

\clearpage 

\begin{deluxetable}{lccc}
\tablecolumns{5}
\tablewidth{0pt}
\tablecaption{NS Masses and Radii from Model Fits for 4U 1728-34\tablenotemark{b}.}
\tablehead{\colhead{Distance}&\colhead{m}&\colhead{$R_{NS}$}&\colhead{$\chi_{red}^2$}\\ \colhead{kpc}&\colhead{$M_\odot$}&\colhead{km}&\colhead{$\chi^2/$DoF}}
\startdata
4.00 &$0.91 \pm 0.02 $&$8.66^{+0.08}_{-0.10}$& 0.963 \\
4.25 &$1.06^{+0.03}_{-0.02}$&$8.92^{+0.10}_{-0.13}$& 0.963 \\
4.50 &$1.25^{+0.06}_{-0.04}$&$9.00^{+0.17}_{-0.28}$& 0.963 \\
4.75 &$1.48 \pm 0.04$&$8.86\pm0.23$& 0.972 \\
5.00 &$1.61 \pm 0.02$&$9.60^{+0.12}_{-0.11}$& 1.102 \\
\enddata
\tablenotetext{a}{errors are given for 90\% of confidence}
\end{deluxetable}

\newpage

\newpage
\begin{figure}
\includegraphics[width=5in,height=6in,angle=-90]{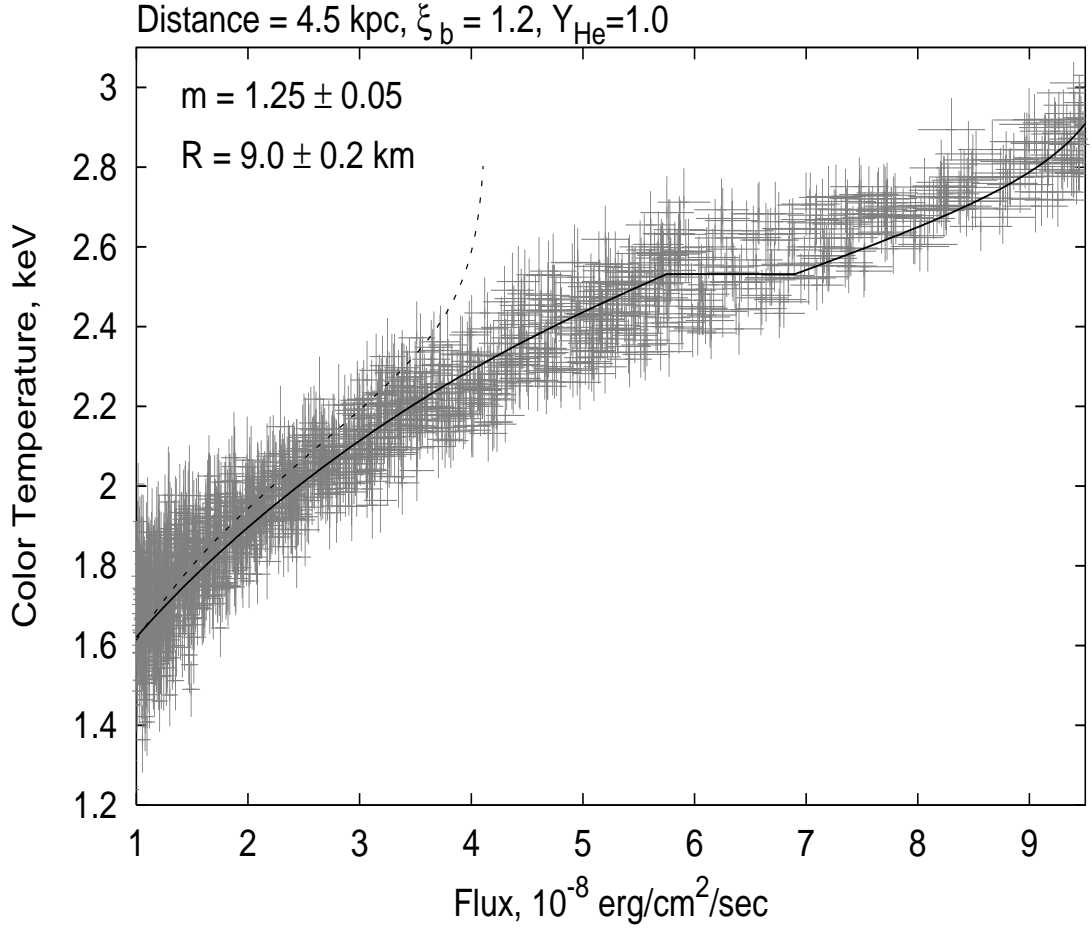}
\caption{Color blackbody  temperature of burst spectra for 26 bursts from
 4U 1728-34 versus flux. Solid curve presents the analytical model 
fit with fixed $d_{10}=0.45$ and $Y_{He}=1.0$. Dashed line present 
the same model with $Y_{He}=0$.\label{fit} }
\end{figure}

\newpage
\begin{figure}
\includegraphics[width=5in,height=7.in,angle=-90]{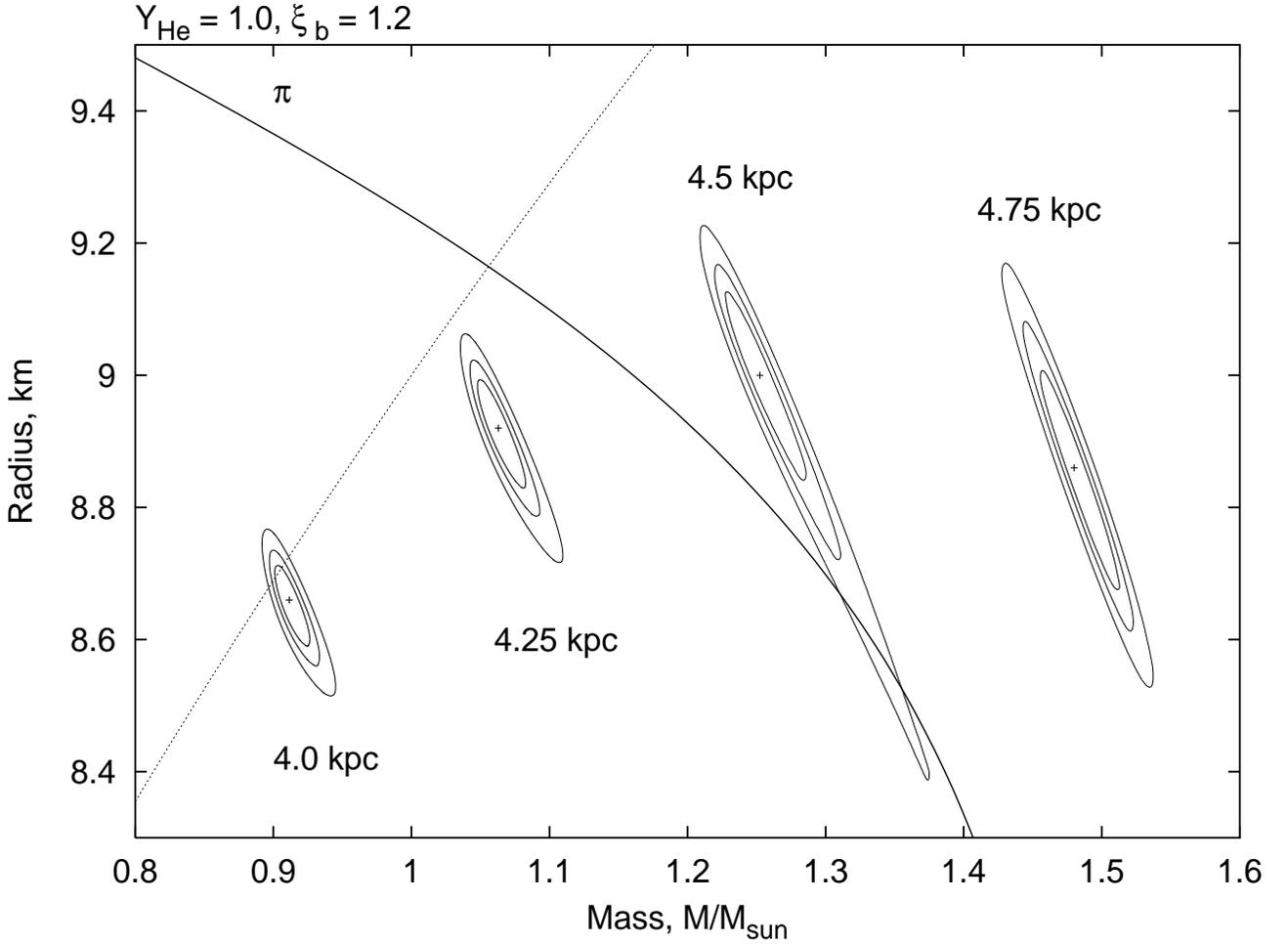}
\caption{ Mass-radius contour obtained by the model fitting. 
Mass-radius relations  for soft EOS \citep{bp} is presented by solid line. 
The dotted curve shows the dependence of the accretion disk inner edge $R_{in}$ obtained using
 the transition layer model (TLM).\label{contours}}
\end{figure}

\end{document}